# Polarized Neutrons at ISIS: Recent Developments And Highlights

Polly MITCHELL[1,§] and Holly I. BARNFIELD[1],
Mark DEVONPORT[1], Kirill NEMKOVSKI[1*], Gøran J. NILSEN[1*],
Peter GALSWORTHY[2], Gavin B. G. STENNING[3]

[1]*Neutron and Muon Instrumentation Development Group, ISIS Neutron and Muon Source, STFC Rutherford Appleton Laboratory, Harwell Science and Innovation Campus, Didcot, Oxfordshire, OX11 0QX, UK*
[2]*Design Division, ISIS Neutron and Muon Source, STFC Rutherford Appleton Laboratory, Harwell Science and Innovation Campus, Didcot, Oxfordshire, OX11 0QX, UK*
[3]*Materials Characterisation Laboratory, ISIS Neutron and Muon Source, STFC Rutherford Appleton Laboratory, Harwell Science and Innovation Campus, Didcot, Oxfordshire, OX11 0QX, UK*

*E-mail: *kirill.nemkovskiy@stfc.ac.uk*, *goran.nilsen@stfc.ac.uk*

We present two recent projects which aim to improve the performance of polarized neutron scattering experiments using hyperpolarized $^3$He spin filters at ISIS. The first is the optimization of a new compact magnetostatic cavity ("Magic Box") to house the $^3$He spin filters based on an existing design. With a length of only 380 *mm*, it provides a field gradient relaxation time for the $^3$He cell of 421 *h* in ambient conditions. It also contains a radiofrequency coil for adiabatic fast passage flipping. The second project is dedicated to the improvement of the $^3$He relaxation time inside the spin filter cell. We have developed a chamber which allows for the deposition of alkali metal coatings on the surface of substrates. This emulates the spin filter cell walls, as well as subsequent heat treatment, thus mimicking the preparation of a new spin filter cell. The chamber is air-tight and has transparent windows, so that the structure resulting from the deposition of alkali metal on the surface of the wafer can be studied by X-ray or neutron reflectometry. We plan to continue this work by performing a systematic study at various conditions, which should help to shed light on the long-standing mystery of how alkali metal coatings help to improve relaxation time of $^3$He cells. The first results are discussed in the text.

## 1. Introduction

The use of polarized neutron beams can improve virtually all types of neutron scattering experiment, either by providing more information on the cross section components or by enhancing the resolution of the instrument using spin echo. An attractive technology to either polarize the incident beam or analyze the polarization of the scattered beam is the hyperpolarized $^3$He spin filter. This exploits the dependence of the neutron absorption cross section on the combined spin of the neutron-$^3$He system to select a single spin state from a neutron beam. Advantages of $^3$He spin filters include their relatively wide bandwidth, adaptability for large solid angle applications, and low cost per solid angle relative to supermirror devices.

---

[§]Present address: Max-Planck-Institute for Chemical Physics of Solids, Dresden, Germany

The Polarized Neutrons section at ISIS (POLNS) is a part of the Neutron & Muon Instrument Development group (NMIDG). It is responsible for the development of new polarized instrumentation and software, as well as the provision of polarized infrastructure for several instruments. Typically, the POLNS is involved in the support of experiments which exploit hyperpolarized $^3$He spin filters. The $^3$He-based setup is already well established for the uniaxial neutron polarization analysis at the cold time-of-flight spectrometer LET [1-3], where polarized experiments are becoming routine [4-6]. Currently, we are also expanding the use of $^3$He spin filters for small-angle scattering instruments such as Larmor [7] and Zoom [8], as well as exploring polarized neutron applications for magnetic imaging on the IMAT instrument [9].

There are two main techniques to polarize $^3$He gas [10]. In the spin-exchange optical pumping (SEOP) one uses a mixture of $^3$He gas with alkali-metal vapor. Alkali metal atoms are polarized by direct optical pumping and then transfer the polarization to $^3$He atoms via collisions. This technique deals with bar-scale gas pressures and is compatible with in-situ polarization. As a drawback, the housing for the cell and SEOP infrastructure when using in-situ polarization are relatively bulky and may not fit inside instruments with space limitations. This also reduces the maximum solid angle that can be covered, which is a major limitation when using the device as an analyzer on a wide-angle instrument.

At ISIS, to polarize the $^3$He we use an alternative method, metastable exchange optical pumping (MEOP). The MEOP filling station named FLYNN was installed at ISIS in 2011 [11]. MEOP involves direct pumping by circularly polarized light of the $^3$He metastable $2^3S_1$ state [12] at low gas pressures (~1 mbar), which means that the gas needs to be compressed before it is filled into the spin filter cell and transported to the area of use. In contrast to the in-situ implementation of SEOP, the cell is used separately from the pumping equipment and cell housing tends to be compact. With respect to ex-situ SEOP, the pumping times are much shorter, and the cells can be shaped into any geometry required, as there are fewer limitations on both cell material and because the pressures used tend to be lower, since polarization happens at room temperature. Hence, MEOP is suitable for use in confined spaces, and especially when large solid angle coverage is needed, and also allows a higher throughput of experiments. On the other hand, the polarization of $^3$He in the cell decays with time, reducing both the transmission of neutron beam and its polarization [12]. Thus, when using MEOP, the spin-filter cell must regularly be replaced by a newly polarized one. To minimize the number of cell replacements during the experiment, it is therefore necessary to increase the lifetime of the $^3$He polarization inside the cell as much as possible.

The decay of $^3$He polarization is usually measured by nuclear magnetic resonance (NMR). So, the lifetime is typically expressed in terms of the NMR longitudinal relaxation time $T_1$. There are three contributions to the cell relaxation time [13]: dipole-dipole interaction $T_1^{dd}$ [14], wall relaxation $T_1^w$ [15], magnetic field gradient $T_1^{fg}$ [16].

The total relaxation time is given by

$$\frac{1}{T_1} = \frac{1}{T_1^{dd}} + \frac{1}{T_1^{w}} + \frac{1}{T_1^{fg}} \qquad (1)$$

The dipole-dipole relaxation term is proportional to $^3$He pressure [14]:

$$\frac{1}{T_1^{dd}} = \frac{p}{807} \text{ (h}^{-1}) \qquad (2),$$

where p is the pressure in bar, and the lifetime is measured in hours. For neutron beam experiments, the $^3$He pressure is determined using standard expressions given the requirements of neutron polarization and transmission (see e.g. [2]) for the wavelength range of interest. For cold neutrons (as used on the ISIS instruments listed above), the product of pressure per cell length (so called pressure length) is usually less than 10 *bar·cm*, which for a good field environment and typical cell length of 10 *cm* makes the dipole-dipole term the smallest in equation (1).

An important part of our research-and-development activities is increasing the relaxation time of our $^3$He cells. In this paper we present two recent projects aiming to do it by the reduction of the last two terms in equation (1). The first one is dedicated to the design of a new compact magnetostatic cavity for $^3$He spin-filter cells (so-called "Magic Box"), which is used to minimize the magnetic field gradient term, the contribution of which to the relaxation time is [16]:

$$\frac{1}{T_1^{fg}} = \frac{6700}{p} \left|\frac{\nabla B_\perp}{B}\right|^2 \text{ (h}^{-1}) \qquad (3),$$

where $\left|\frac{\nabla B_\perp}{B}\right|$ is the normalized transverse field gradient averaged over the volume of the $^3$He cell, in units of *cm$^{-1}$*. The second project is dedicated to the study of (with the future aim of minimizing) the second term in equation (1), namely the $^3$He relaxation due to the interaction with cell walls $1/T_1^w$. Historically, the most common material for the $^3$He cells was glass, and in particular aluminosilicate GE180 for neutron scattering applications [17]. However, the glass structure strongly depends on the preparation method. Besides, it is hard to entirely exclude the impurities which are harmful to the $^3$He polarization (e.g. Fe). Therefore, it is difficult to reproducibly manufacture high-quality cells [18-22]. Due to this, it is desirable to explore more "regular" materials – for example, amorphous $SiO_2$ ("quartz"), which can be produced with high chemical purity [23] and, especially, single crystalline silicon. Cells made completely from silicon have been recently manufactured and are being tested at the ILL [24].

An essential measure to ensure a large relaxation time of $^3$He inside the cell is to deposit a thin layer of alkali metal – usually caesium or rubidium - on its inner surface [17]. This is conventionally done by distillation but can also be achieved by pipetting molten high-purity metal inside the cell. Then the cell is "baked" at a moderately high temperature (about 100ºC) to coat the inner surfaces.

Alkali metals have successfully been used to improve cells relaxation time for decades. However, the exact mechanisms of the underlying processes are still poorly understood, with a commonly discussed reason being the low energy of adsorption for $^3$He on alkali metals [25]. In this paper we present, to the best of our knowledge, the first attempt at a systematic study of the surface properties of $^3$He cells, aiming to reveal what structure the alkali metal (here caesium) forms on the silicon substrate, and how its physical and chemical properties are affected by the baking temperature and time.

## 2. New compact Magic Box

A highly homogeneous magnetic field in the volume occupied by $^3$He spin filter can be realized in different ways, depending on the geometry and size of the instrument, as well as the shape of the spin filter cell itself. For example, at LET, with its limited space around the sample and large solid angle banana detector (and, correspondingly, $^3$He cell), the magnetic field is provided by a set of compensated Helmholtz coils [2]. On the other hand, for small-angle neutron scattering (e.g. Zoom) and other experiments with similar geometry (e.g. spin echo on Larmor), a good solution can be the use of an elongated magnetostatic cavity, either based on a solenoid for field directions parallel to the beam or a rectangular cavity for directions perpendicular [26]. Typically, the walls of the latter type of cavity are made from mu-metal to both reduce the magnetic field gradients and shield from magnetic interference, with coils wound on the side walls to provide a constant magnetic field. The cavities are also usually equipped with a perpendicular coil to perform adiabatic fast passage (AFP) flipping of the $^3$He polarization [27], which is conceptually similar to the adiabatic resonance radiofrequency (RF) flipper for neutrons [28]. Within the Magic Box, the AFP cell replaces a separate flipper for the neutron beam in the experimental layout, saving space. AFP flipping is also generally near 100% efficient [29]. A general layout of a Magic Box with $^3$He spin filter inside the static coils is shown in Fig.1a.

At the outset of the project, the main type of Magic Box used at ISIS was an early design made at the ILL. Although its performance was still sufficient in most circumstances, for some experiments with stricter space constraints its length of 690 mm was too large. Thus, the aim was to redesign the current Magic Box to make it more compact and easier to use on the beamline, while preserving the homogeneity of the field within. This was to be achieved using electromagnetic simulations to optimize the field gradient inside the box for a variety of different parameters and dimensions.

Some preliminary calculations towards this goal were performed using the Radia package [30]. However, Radia does not support time dependence in its modelling of the field and was thus unsuitable for modelling the AFP flipper coil. For the full modelling of the Magic Box, we therefore switched to the CST studio software suite [31]. The latter allows to include not only magnetostatic but also time-dependent fields in the model. The Radia results were used for extra verification of static field components in the CST model. We also checked (both analytically and via CST simulations) that the self-inductance effects produced by the RF coil do not distort the AFP pulse for the parameters (current, frequency, and frequency sweep rate) well in excess of those typically used in the device. All calculations were carried out on a conventional desktop computer (2.9GHz processor, 16Gb RAM).

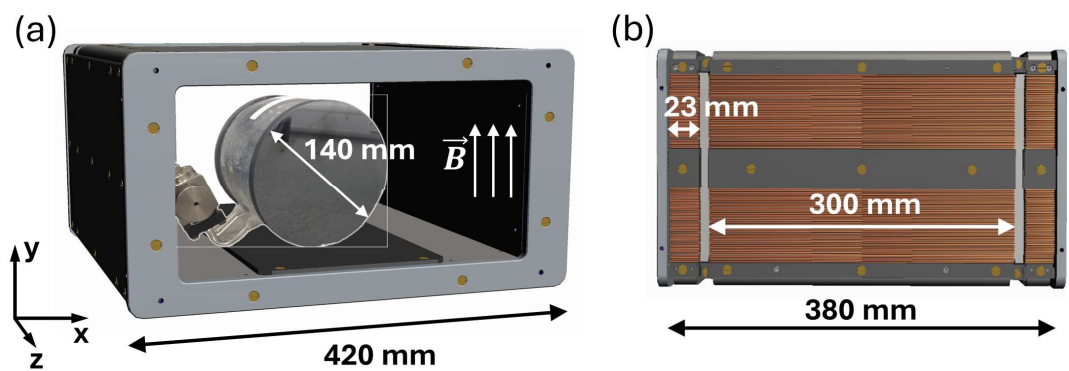

**Fig. 1.** a) General view of the Magic Box with an Si-windowed $^3$He spin filter cell inside. White arrows show the direction of the magnetic field. b) Side view showing sections for main coils (middle) and compensation coils (ends) which run at a higher current. All dimensions are shown for the final optimized configuration (see text).

The starting point for the optimization was a more recent ILL design of a compact Magic Box, the drawings of which were provided by D. Jullien [24]. Its outer dimensions are 420×225×496 $mm^3$ (width × height × length). The walls of the box are made from mu-metal with a thickness of 2 $mm$, with two layers of mu-metal sheets on the sides with the static field coils wrapped around the sheets. The box is held together with a polyethylene carbonate frame.

Unlike the previous Magic Box, the new design also has end-compensation coils (see Fig.1b), similar to those described in [17]. They reduce "end effects" caused by the open faces of the box and maintain the homogeneous field even with a relatively short device length.

The largest $^3$He cell to be used in the combination with new Magic Box has a cylindrical shape with a diameter of 140 $mm$ and length of 100 $mm$ (see Fig.1a). Thus, the box has to provide a homogeneous magnetic field within this volume.

A few test simulations allowed us to observe that the height and the width of the box did not strongly affect the field gradient. And it is the length rather than the width and the height that is the main spatial constraint on the Larmor and Zoom instruments. On the other hand, the box dimensions had to be sufficient for the cell to be lifted in and out of it with ease. Therefore, we decided to remain with the ILL box height and width, while the optimal length had to be determined via simulation.

The remaining main parameters to optimize were then the box length, the size and position of the gap between the main and compensation coils, and the ratio between their currents $I_{comp}/I_{main}$. As CST Studio does not support optimization with custom figures of merit, a parameter sweep was undertaken for the current ratio, gap position and gap width, at nine length values between 310 $mm$ to 450 $mm$. Then, for each length, we determined the set of parameters which provide the minimum value of the normalized transverse field gradient from formula (3) averaged over the cell volume.

The aforementioned parameters are correlated, and independent variation of all three of them across a broad range (i.e. a grid scan) is not feasible. Our approach instead focused on tuning the current ratio, as it has the strongest influence on the field gradient near the optimal configuration. Based on the original ILL design, we began with a geometry in which the main coil occupies approximately 75% of the box length, 10% of the length is allocated to two compensation coils, and remaining space is reserved for the gaps in between. For a given geometry we determined the optimal current ratio, and then iteratively adjusted the gap width and position to further refine the field gradient. This process was repeated several times to ensure that no further improvements could be made to the field gradient through adjustments to these parameters.

The optimal current ratios $I_{comp}/I_{main}$ for different box lengths are shown in Fig.2a as dotted pink diamonds (right axis). For our configuration, the optimized ratio values fall in the range between 1.45 (for the shortest length) and 1.9 (the longest). The increase of the required $I_{comp}/I_{main}$ ratio along with box length reflects the decreasing importance of the compensation coils for longer boxes.

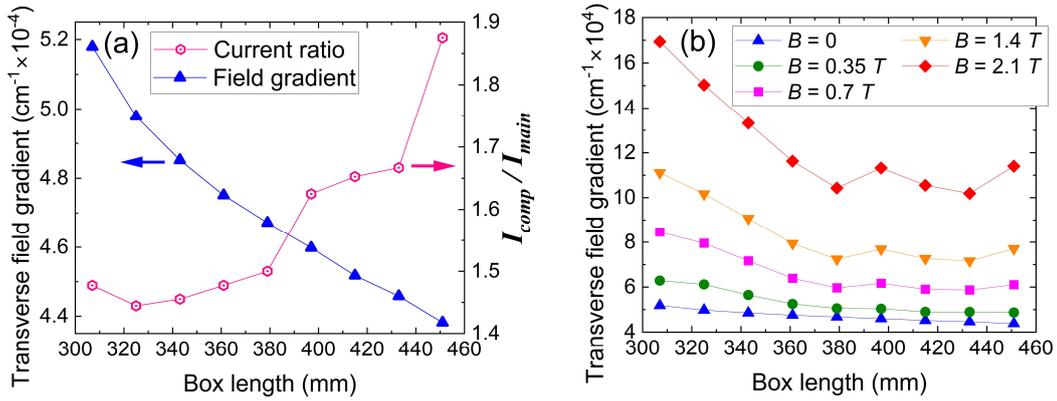

**Fig. 2.** Results of the CST simulations. a) Pink dotted diamonds (right axis) - optimal current ratio $I_{comp}/I_{main}$ in the static coils as a function of the box length (gap width and position are set to their optimized numbers for each particular length value). Blue up-triangles (left axis) - normalized volume-averaged transverse magnetic field gradient taken within a volume of $140 \times 140 \times 100$ mm$^3$ (width × height × length) at the box center for a range of box lengths. b) Field gradient in the presence of the HTS magnet for several values of the field inside the magnet. The magnet is placed 100 mm from the front box face. The direction of magnetic field in the magnet is parallel to the box.

The transverse field gradient is presented in Fig.2a as blue up-triangles (left axis). As expected, the gradient value is higher (i.e. worse) as the box length decreases. In the range of length between 450 and 360 *mm*, the dependence is nearly linear, while below 360 *mm* the increase becomes steeper. Therefore, we consider 360 *mm* as the minimum box length with acceptable gradients. (Here and below, we consider the gradient of the static magnetic field only. The gradient of the RF field was found to have similar dependence on the box length; however, its contribution is about two orders of magnitude weaker than a static one. Therefore, in the presented analysis we neglect it.)

In some experiments the new Magic Box will be used in combination with a magnet at the sample position. The typical magnet for this purpose at the Larmor and Zoom instruments is the High Temperature Superconducting (HTS) 3*T* horizontal magnet [32]. Hence, we have investigated how the presence of this magnet will affect the homogeneity of the magnetic field within the Magic Box. The worst-case scenario corresponds to the configuration of the magnet with field produced parallel to the neutron beam and the box 100 *mm* from the external exit face of the magnet. In this configuration we have performed measurements of the magnetic field distribution outside the magnet within the volume where the magic box would sit using the same method above. The resulting field was modelled in CST using a conventional coil, then added to the simulation of the magic box.

The resulting transverse field gradients are shown in Fig.2b. For all values under consideration, the field gradient inside the box weakly depends on the box length down to 380 *mm*. However, for shorter box lengths the gradient increases drastically, especially for higher values of the external field. Thus, the length of 380 *mm* seems to be a good compromise between box compactness and performance.

Based on the above considerations, for our final design we have chosen the following parameters:

- Box length: 380 *mm*
- Main coil length: 300 *mm*
- Compensation coil length: 23 *mm*
- Gap width: 17 *mm*
- Current ratio $I_{comp}/I_{main}$: 1.5

The coils are placed with their middle positions at ¼ and ¾ of the height of the cell (see Fig.1b). On each side there are two coils, with 49 turns for each coil.

The final optimized volume-averaged transverse field gradient is $4.67\times10^{-4}$ cm$^{-1}$, which corresponds to the field-gradient contribution to the relaxation time of $T_1^{fg} = 615\ h$ for a cell pressure of 0.9 *bar*.

In the presence of the HTS magnet with inductance of B = 2.1 *T* at the sample position, we get the field gradient in the box of $10.4\times10^{-4}\ cm^{-1}$ and field-gradient relaxation time of $T_1^{fg} = 124\ h$.

## 3. Study of caesium deposition on single-crystalline silicon

The main challenge of studying the properties of alkali metal coatings inside real $^3$He cells in situ is the difficulty of accessing the inner surface of the cells given the high reactivity of the caesium. Therefore, we attempt to emulate the same conditions in a model system consisting of a single crystalline silicon wafer inside a chamber that contains a slot for the wafer, a reservoir for caesium coating (located underneath the wafer), and suitable windows for X-ray and neutron reflectometry studies (see Fig.3). The main body of the chamber is made from aluminium. The silicon wafer is positioned in the bottom half of the assembly. Under the wafer there is a reservoir for caesium metal with four gaps for caesium vapor to exit. For the reflectometry experiments, the chamber is equipped with removable windows with air-tight clamps, while for caesium deposition and baking, the top section is replaced by a blank aluminium section to avoid compromising the windows at high temperature. For X-ray experiments the windows are Kapton foil. For future neutron experiments, the Kapton can be replaced by aluminium.

The wafers were ordered from an industrial manufacturer (Inseto [33]). They were of 50 *mm* diameter and 0.5 *mm* thickness, single-side polished, with the (111) crystallographic axis perpendicular to the surface. The caesium was obtained from Sigma Aldrich (99.95% purity) and was deposited in the reservoir by pipette inside an argon glovebox. To mimic the procedure of cell baking, the whole chamber (with blank top section) was loaded into an oven and heated at various temperatures up to 150ºC for up to 1 week and then cooled down to the room temperature. For each baking stage the top section with windows was interchanged with the blank one.

X-ray reflectometry (XRR) experiments were carried using the Rigaku SmartLab reflectometer at the ISIS Materials Characterisation Laboratory with wavelength λ=1.541 *Å* (Cu Kα). Fitting of the obtained reflectometry curves was performed using the GenX software [34].

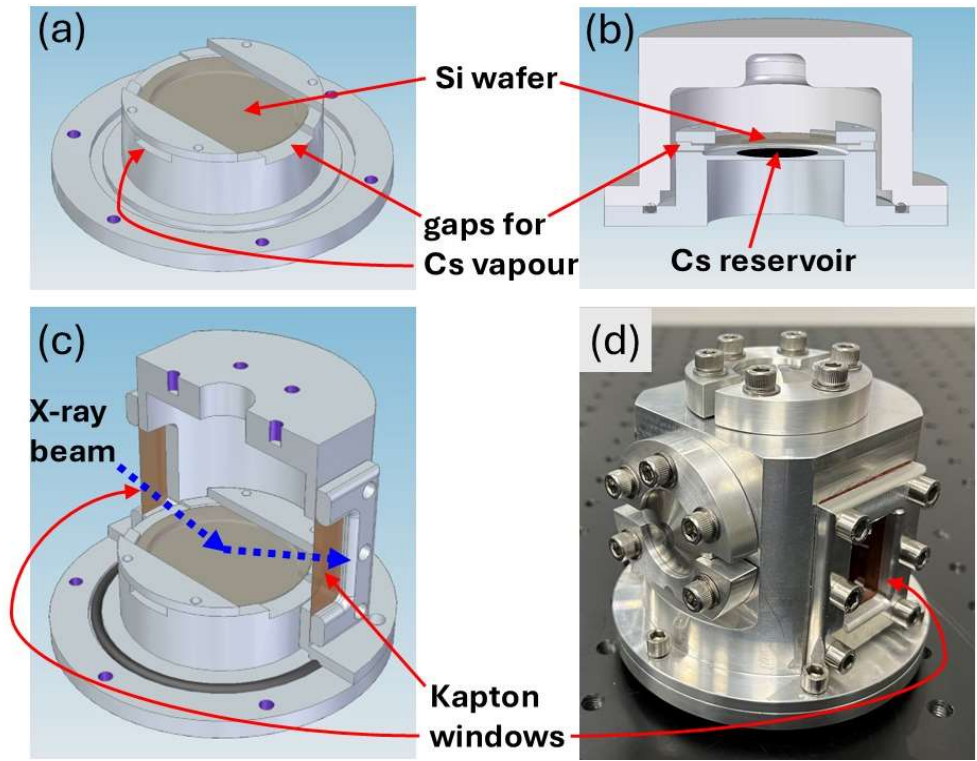

**Fig. 3.** The chamber for the reflectometry studies of caesium deposition on the silicon wafer. a) Basement with mounted Si wafer. b) Cut view of the assembly with top section for caesium deposition and baking procedure. c) Cut view of the assembly with top section for X-ray reflectometry measurements. d) Photo of the assembly with top part for X-ray reflectometry measurements. Caps screwed on the top and the left side cover the holes for pumping the chamber out and filling by inert gas (not used in the current experiment).

Fig.4 presents the first experimental results obtained using the chamber with the wafer prepared under various conditions. Initially, without caesium, the wafer exhibits a typical fringe corresponding to an amorphous silicon oxide surface layer. When caesium is added without baking, the XRR curve decays more steeply, indicating higher surface roughness (here and below we refer to the Root Means Square (RMS) roughness). Given that the caesium is introduced at a temperature only slightly over the melting point, it is surprising that such a strong effect is observed.

Baking the chamber at $50°C$ for 12 hours causes only a slight increase in the steepness of the decay, with minimal impact to the profile. However, after baking at 100ºC for one week, the profile shows additional features compared to the initial wafer. The profile plateaus at a lower intensity and exhibits an interference peak around $Q = 0.25$ $Å^{-1}$. After baking at 150ºC for one week, an extra peak appears at $Q = 0.55$ $Å^{-1}$. More exact information can be obtained by the fitting the observed curves to a slab model containing the substrate, the $SiO_2$ surface layer, and (where relevant) a nominal Cs layer on top. All parameters aside from the substrate scattering length density were allowed to vary. The results of the fitting are summarized in Table I.

The plain silicon wafer (no Cs) has a surface layer of silicon dioxide with a thickness of $19.8±0.6$ $Å$ and an RMS roughness of $3.6±0.1$ $Å$. In the presence of caesium in the chamber, the characteristics of the $SiO_2$ layer remain nearly the same. However, upon baking, SiO$_2$ starts to interact with Cs (see below), and this reduces the layer thickness, as well the roughness. After the initial deposition without baking, the thickness of the Cs layer (15.8 Å) exceeds the lattice parameter of solid Cs (6.14 Å) by approximately 2.5 times, while the density of the layer is about one half of the one for the solid Cs. This means that Cs layer is either not (poly)crystalline or that the coverage is inhomogeneous. Baking at 50ºC gives rise to a decrease of Cs layer thickness along with increase of the density. The general trends upon heating and time (which is more important remains a question to address with future experiments) are the reduction and smoothing of the SiO$_2$ interface layer and further growing of Cs layer. After baking at 100ºC Cs layer can be distinguished visually.

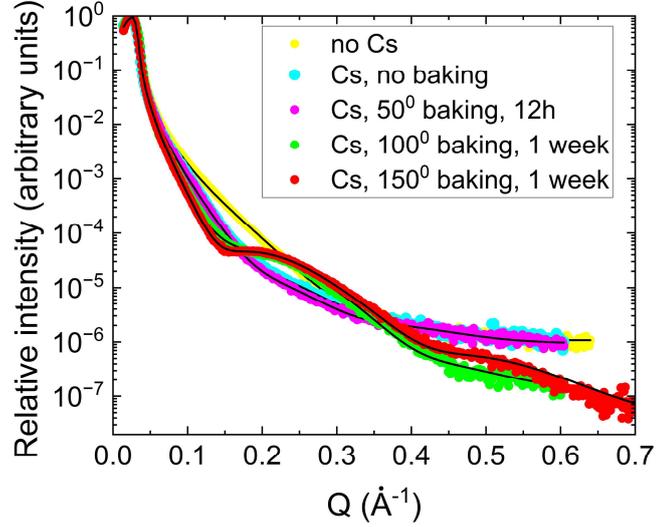

**Fig. 4.** XRR profiles measured on wafer without caesium, coated by caesium without additional baking, and with heating at 50ºC during 12h, 100ºC during one week and 150ºC for one week. Symbols – experiment, lines – fitting (see text).

For the measurements after the baking at higher temperatures (100ºC and 150ºC), the fit allows to describe main features of the XRR patterns, but the values of $\chi^2$ are considerably higher, especially for 150ºC where $\chi^2$ is a one order of magnitude higher (see Table I). The reduction of the fitting quality may reflect the fact, that our original fitting model (Si substrate, SiO2 interface and Cs on the top) becomes less and less adequate due to the interaction between silicon wafer and Cs coating.

In conclusion, we wish to note that the fit parameters for the 100ºC data fall outside the general trend observed between the other dataset. Although this may result from a technical issue with the measurement, several other explanations, including surface inhomogeneities and temperature- and time-dependent chemical transformations are also possible.

**Table I.** Fitting results for XRR profiles. All measurements have been performed at room temperature. Temperature values in the first column refers to the baking conditions.

|  | SiO$_2$ layer | | | Cs layer | | | $\chi^2$ |
|---|---|---|---|---|---|---|---|
|  | thickness (Å) | roughness (Å) | density ×10$^4$ (F.U./Å$^3$) | thickness (Å) | roughness (Å) | Density ×10$^4$ (F.U./Å$^3$) |  |
| no Cs | 19.8±0.6 | 3.5 ± 0.1 | 270 ± 3 | – | – | – | 2.7 |
| Cs, no baking | 19.5±0.5 | 3.4 ± 0.1 | 276 ± 5 | 15.8 ± 0.3 | 3.3 ± 0.3 | 42 ± 2 | 3.9 |
| Cs, 50º | 19.2±0.6 | 2.6 ± 0.1 | 269 ± 3 | 14.3 ± 0.3 | 5.3 ± 0.3 | 57 ± 2 | 2.8 |
| Cs, 100º | 8.3 ± 0.5 | 3.0 ± 0.2 | 270 ± 2 | 21.9 ± 0.3 | 4.0 ± 0.3 | 37 ± 1 | 9.5 |
| Cs, 150º | 16 ± 2 | 1.6 ± 0.2 | 266 ± 5 | 21.1 ± 0.5 | 3.8 ± 0.7 | 53 ± 3 | 157 |

## 4. Summary

We have designed and manufactured a compact Magic Box with a length of 380 *mm*. This is 43% shorter than our previous magnetostatic cavity (and 25% shorter than a more modern end-compensated design) but is expected to have similar or even better performance, including in the presence of a high-$T_c$ superconducting magnet at the sample position. Preliminary tests show that in laboratory conditions (in the absence of magnet) a relaxation time of 130 *h* at the $^3$He pressure of 0.9 *bar* is obtained for a cell with a wall relaxation time of 238 *h*. We thus estimate the field-gradient component to be at least 421 *h* (that corresponds to an average normalized field gradient of $5.65 \cdot 10^{-4}$ cm$^{-1}$). Although this number is high enough for most our practical applications, the deviation from the predicted 615 *h* leaves certain room for further improvements (in particular, we consider rewinding the coils) and optimization of operational currents.

Also, we have started a systematic investigation of the chemical and physical surface modifications caused by caesium deposition on a single crystalline silicon substrate intended to model the wall of the $^3$He cell. To reproduce the conditions inside the cell we have developed a dedicated chamber which allows to deposit caesium on the silicon wafers, and which has windows for X-ray or neutron reflectometry measurements. First results show the evolution of caesium layer upon heating at different temperatures. The most intriguing is the result obtained at the highest temperature in the range (150ºC) where we see clear indications that caesium does not form a simple metallic layer but instead interacts chemically with the oxide layer on the substrate surface. More detailed and systematic investigations, which will include the repetition of the same XRR measurements on the series of identical Si wafers, are ongoing, with neutron reflectometry studies also planned for the future.

**Acknowledgment**

We would like to thank David Jullien (ILL) for numerous discussions and invaluable help in both of the projects discussed above, and in particular for the provision of his design of the Magic Box we used as a starting point. We also grateful to Sam Pitman (ISIS) for the consulting on CST software, Daniel Nye (ISIS) for the assistance in the reflectometry experiments, Christy Kinane and Andrew Caruana (ISIS) and Artur Glavic (PSI) for the help with analysis of the reflectometry data, Duc Le (ISIS) for the help in testing new Magic Box, and Steve Parnell, Diego Alba Venero, Dirk Honecker, Andy Church and Oleksandr Tomchuk (ISIS) for fruitful discussions. We also highly appreciate the communication with Holly McPhillips (University of Kent, Canterbury) who was working on a related project for $^3$He cells before (not included in this paper) and kindly shared her experience.